\documentclass[a4paper,11pt]{article}
\pdfoutput=1 

\usepackage{jinstpub} 

\usepackage[version=3]{mhchem}
\usepackage{siunitx}
\usepackage{natbib}
\usepackage{amsmath} 

\usepackage{verbatim}

\usepackage{setspace}
\usepackage{graphicx}
\usepackage{amsmath}
\usepackage{siunitx}

\usepackage{pbox}
\usepackage{xcolor}

\newcolumntype{"}{@{\hskip\tabcolsep\vrule width 1pt\hskip\tabcolsep}}
\DeclareSIUnit[number-unit-product = \;]\year{yr}
\DeclareSIUnit\parsec{pc}
\DeclareSIUnit\torr{Torr}
\DeclareSIUnit\centimeter{\centi \meter}
\DeclareSIUnit\sq{\ensuremath{\Box}}
\usepackage{textcomp}
\usepackage{caption}
\usepackage{subcaption}
\usepackage{url}
\usepackage{ctable}

\title{Gas Gains Over 10$^4$ and Optimisation using $^{55}$Fe X-rays in Low Pressure SF$_6$ with a Novel Multi-Mesh ThGEM for Directional Dark Matter Searches}

\author[a,1]{A.G. McLean,\note{Corresponding author.}}
\author[a]{N.J.C. Spooner,}
\author[b]{T. Crane,}
\author[a]{C. Eldridge,}
\author[a]{A.C. Ezeribe,}
\author[a]{R.R. Marcelo Gregorio,}
\author[a]{and A. Scarff}

\affiliation[a]{Department of Physics and Astronomy, University of Sheffield, South Yorkshire, S3 7RH, United Kingdom}
\affiliation[b]{AWE plc, Aldermaston, Reading, Berkshire, RG7 4PR, United Kingdom}

\emailAdd{ali.mclean@sheffield.ac.uk}

\abstract{The Negative Ion Drift (NID) gas SF$_6$ has favourable properties for track reconstruction in directional Dark Matter (DM) searches utilising low pressure gaseous Time Projection Chambers (TPCs). However, the electronegative nature of the gas means that it is more difficult to achieve significant gas gains with regular Thick Gaseous Electron Multipliers (ThGEMs). Typically, the maximum attainable gas gain in SF$_6$ and other Negative Ion (NI) gas mixtures, previously achieved with an $^{55}$Fe X-ray source or electron beam, is on the order of $10^3$\cite{Phan2017, Baracchini2018, Miyamoto2004, Ishiura2018}; whereas electron drift gases like CF$_4$ and similar mixtures are readily capable of reaching gas gains on the order of $10^4$ or greater \cite{Burns2017, Callum_thesis, Isobe, Ounalli, Marafini2018}. In this paper, a novel two stage Multi-Mesh ThGEM (MMThGEM) structure is presented. The MMThGEM was used to amplify charge liberated by an $^{55}$Fe X-ray source in 40 Torr of SF$_6$. By expanding on previously demonstrated results \cite{Eldridge}, the device was pushed to its sparking limit and stable gas gains up to $\sim$50000 were observed. The device was further optimised by varying the field strengths of both the collection and transfer regions in isolation. Following this optimisation procedure, the device was able to produce a maximum stable gas gain of $\sim$90000. These results demonstrate an order of magnitude improvement in gain with the NID gas over previously reported values and ultimately benefits the sensitivity of a NITPC to low energy recoils in the context of a directional DM search.}


\begin{document}
\maketitle
\flushbottom

\section{Introduction}
\label{sec:intro}
The Multi-Mesh Thick Gaseous Electron Multiplier (MMThGEM) presented in this paper is a novel multistage design variation on the original single stage MMThGEM \cite{rui}. It was designed as a gain stage device for coupling to a micromegas readout plane in a low pressure Negative Ion (NI) gas Time Projection Chamber (NITPC). Such experiments are used in directional searches for Weakly Interacting Massive Particles (WIMPs). 

WIMPs are a hypothetical candidate particle for Dark Matter (DM) which constitutes $85\%$ of the mass in the known Universe. Attempts to detect rare events in which WIMPs elastically scatter off nuclei have made significant improvements in sensitivity in recent years. However, the WIMP-nucleon cross section limits produced by leading two-phase xenon TPC experiments, like those of the LZ and XENON collaborations \cite{LZ,XENON}, are approaching the neutrino fog \cite{OHare2021}. In this region of sensitivity, these detectors are expected to measure Nuclear Recoils (NRs) induced by neutrinos predominately coming from the sun \cite{Billard2014}. With this current leading technology, these recoils will make the positive identification of a NR induced by a WIMP significantly more difficult. 

Low pressure gaseous NITPCs are seen as a viable method for probing into the neutrino fog because the reconstruction of ionisation tracks, produced by NRs in such a target medium, can be used to determine the direction of the incoming particle, resulting in a so-called galactic signature \cite{Vahsen2020}. This is advantageous when trying to discriminate between neutrinos originating from the Sun and WIMP signals which, due to the motion of the Solar System around the Galaxy, are expected to originate from the direction of the Cygnus constellation \cite{Spergel1988, Vahsen2019}. Unlike the DAMA/LIBRA experiment which seeks a galactic signature via an annual modulation of events \cite{dama2018}, which has already been ruled out by more sensitive experiments \cite{Akerib2017, Aprile2017, Agnese2014} and tightly constrained by similar NaI target experiments like COSINE-100 \cite{cosine2019}, a directional galactic signal can not be mimicked by terrestrial background events \cite{Klinger2015, Kudryavtsev2010, Davis2014}. 

The direction of a recoiling nucleus can be reconstructed from the track of ionisation it leaves behind in the gas by utilising a readout plane with positional sensitivity. The DRIFT experiment pioneered this method using back-to-back Multi-Wire Proportional Counters (MWPCs), filled with the NI gas CS$_2$ around 40 Torr, and so far has the best published sensitivity for directional DM searches \cite{Bat2016}. NI gases, like CS$_2$ and SF$_6$, are preferred over electron drift gases like CF$_4$ because they exhibit significantly less diffusion during the drift phase. However, the NI nature of these gases means that it is more difficult to achieve significant gas gains \cite{Phan2017}. For instance, gas gains achieved in CS$_2$ and SF$_6$ typically have an order of magnitude $\leq$ 10$^3$ compared to $\geq$ 10$^4$ in CF$_4$ \cite{Phan2017, Baracchini2018, Miyamoto2004, Ishiura2018, Burns2017, Marafini2018,Isobe, Ounalli, Callum_thesis}. This is likely due to the requirement of the electron to be stripped from the NI before amplification can occur \cite{Phan2017}. As a result, the sensitivity to low energy recoils in Negative Ion Drift (NID) gases is limited by at least one order of magnitude. One exception to this was demonstrated with a triple Gaseous Electron Multiplier (GEM) setup in 100 Torr of SF$_6$ which produced maximum gas gains around $\sim$10000, however the vast majority of measurements achieved with that setup were on the order of 10$^3$ or lower \cite{Ishiura2018} and ideally the pressure should be lower than this in order to elongate potential NR tracks. 

In this paper, we start by introducing the design and operation of the MMThGEM in a low pressure gas. Then the experimental setup and method of measuring the gas gain is discussed along with a calibration run in 40 Torr of CF$_4$. Following this, we expand on previously reported results with the MMThGEM \cite{Eldridge} to push the device to its sparking limit in 40 Torr of SF$_6$. This process reveals that the device is capable of achieving gas gains in SF$_6$ on the order of $10^4$. Further optimisation of the collection and transfer fields in the device demonstrate that this gain can be improved further. Finally, gas gain results are presented comparing the fully optimised device in SF$_6$ to the CF$_4$ calibration run. 

\section{MMThGEM Design and Operating Principles}
\label{sec:design}

\begin{figure} [h]
\captionsetup[subfigure]{justification=centering}
\centering
    \begin{subfigure} {.5\textwidth}
        \centering
        \includegraphics[width=0.7\textwidth]{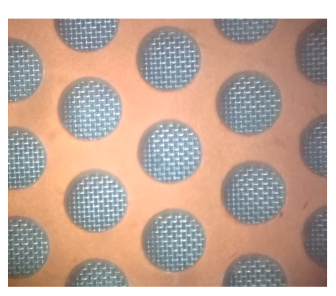}
        \caption{}
        \label{fig:MMThGEM_image}
    \end{subfigure}%
    \begin{subfigure}{.5\textwidth}
        \centering
        \includegraphics[width=\linewidth]{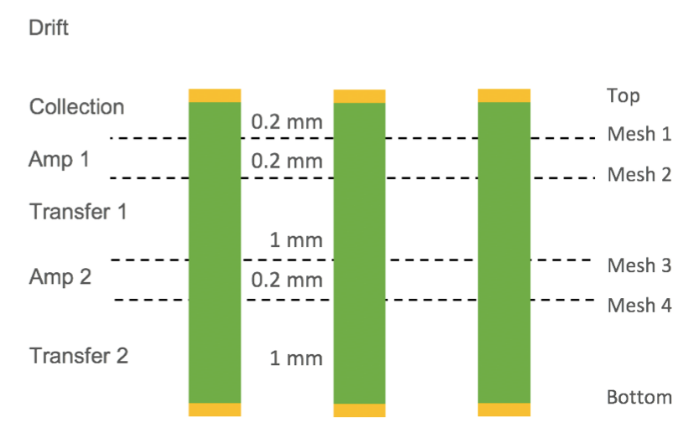}
        \caption{}
        \label{fig:MMThGEM_cross_sec}
    \end{subfigure}
    \caption{(a) Image of the MMThGEM hole structure as viewed from above. (b) Cross sectional diagram of the MMThGEM device. }
\label{fig:MMThGEM_both}
\end{figure}

The multistage MMThGEM presented here was designed and fabricated in 2018 at the micropattern detector production facility at CERN following the initial demonstration of a single stage MMThGEM design \cite{Eldridge,rui}. It was designed as an amplification stage device which could be coupled to a micromegas for use in directional DM searches utilising a NID gas. The device, as shown in figure \ref{fig:MMThGEM_both}, is similar in structure to a regular ThGEM \cite{breskin}. It consists of a dielectric layer between two copper electrode layers with a hexagonal lattice of holes drilled through the device. The addition of four mesh layers embedded in the dielectric layer differentiates the MMThGEM from a regular ThGEM structure. These mesh layers span across the holes and can be seen in figure \ref{fig:MMThGEM_image}.

As shown in the cross sectional diagram of figure \ref{fig:MMThGEM_cross_sec}, the meshes are situated at depths of 200$\mu$m, 400$\mu$m, 1400$\mu$m, and 1600$\mu$m from the top plane of the device. The holes have a diameter of 0.8 mm and a pitch of 1.2 mm. The device has a total thickness of 2600$\mu$m and a total active area of 10 x 10 cm. The addition of the meshes establishes five distinct regions in the detector; the collection field, amplification field 1, transfer field 1, amplification field 2 and transfer field 2. All the layers can be biased individually in order to establish electric fields of varying strengths between neighboring electrodes. As described further in \cite{Callum_thesis}, multi-stage amplification requirements for a low pressure NID gas were motivated by the low gas gains produced by a single ThGEM in SF$_6$. Unlike double/triple ThGEM configurations and other multistage devices like Multi-layer Thick Gaseous Electron Multipliers (MThGEMs) \cite{Cortesi2017}, the meshes in the MMThGEM ensure that the electric fields are uniform and should therefore provide better amplification properties.

When a particle interacts with the gas in the drift region between the top layer and a cathode, the resulting ionised negative charge is drifted towards the MMThGEM under the applied drift field. When the charge reaches the MMThGEM, it is focused into the holes by the collection field applied between the top layer and mesh 1. The charge is accelerated when it reaches the first amplification field due to a large electric field applied between mesh 1 and mesh 2. This acceleration of the charge causes further ionisation through subsequent collisions with gas molecules and causes an avalanche of ionised charge. After the first amplification stage, the first transfer field between mesh 2 and mesh 3 transports the charge to the second amplification stage. The charge is then amplified for a second time between mesh 3 and mesh 4. Under its intended use, this amplified charge would be transported towards a micromegas plane by the second transfer field between mesh 4 and the bottom electrode plane. In the following section, the experimental setup and biasing circuitry is discussed. 

\section{Experimental Setup and Biasing Circuitry}
\label{sec:experiment}

\begin{figure} [h]
\captionsetup[subfigure]{justification=centering}
\centering
\begin{subfigure} {.45\textwidth}
    \centering
  \includegraphics[angle=270, trim ={52cm 22cm 37cm 18cm}, clip, width=\textwidth]{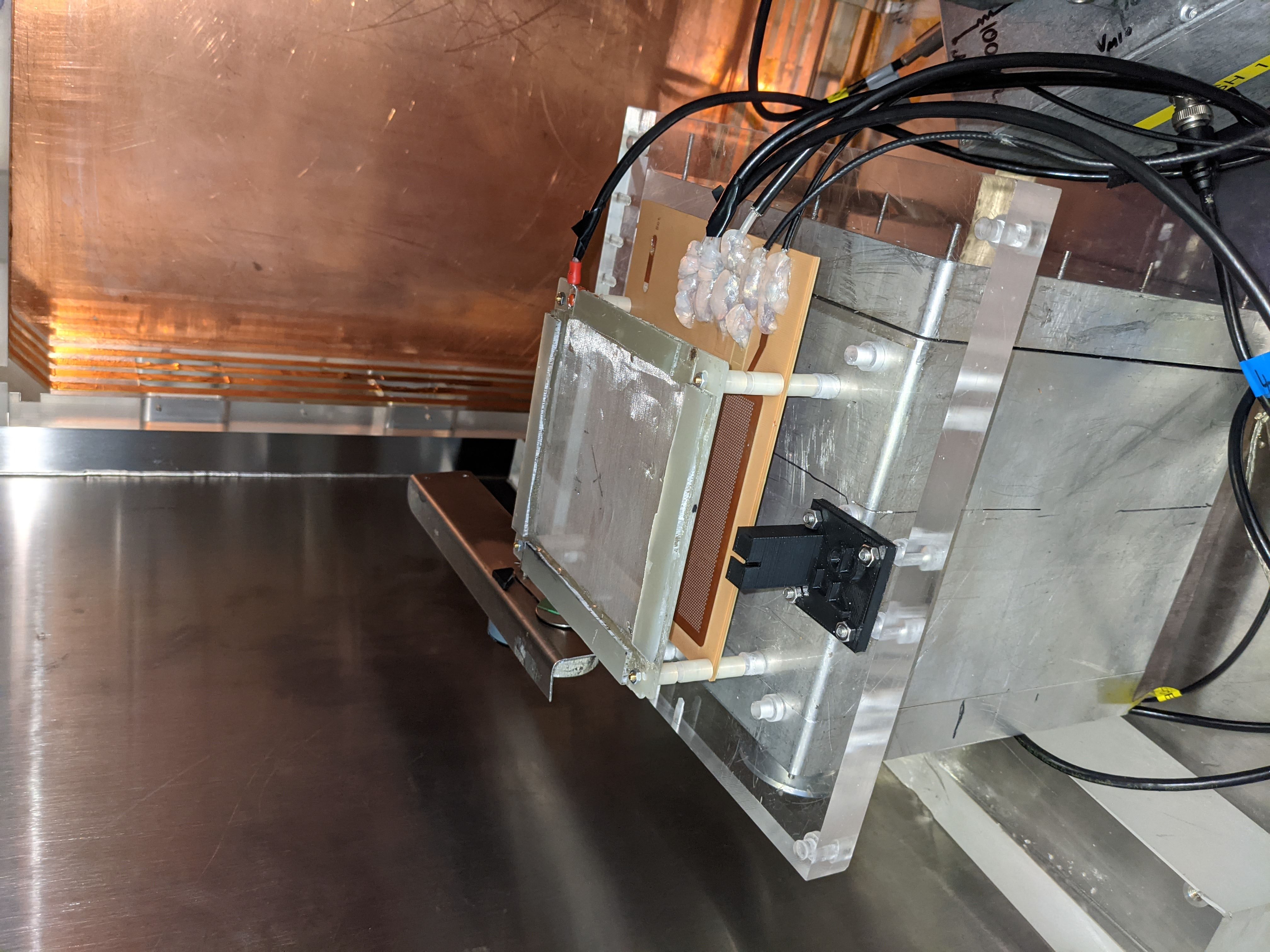}
  \caption{}
  \label{fig:MMThGEM_setup}
\end{subfigure}%
\begin{subfigure} {.55\textwidth}
  \centering
  \includegraphics[trim ={0cm 0cm 2cm 0cm}, clip, width=\textwidth]{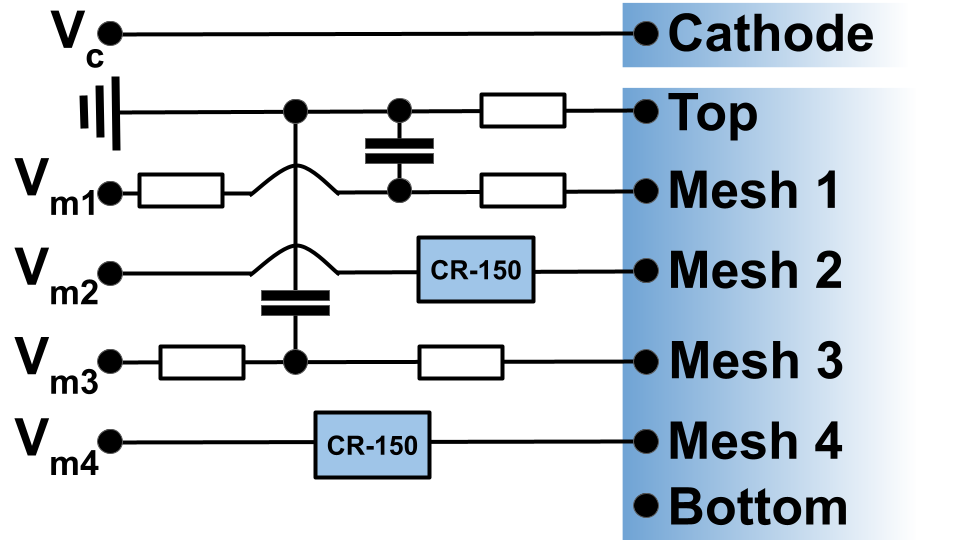}
  \caption{}
  \label{fig:MMThGEM_BIAS}
\end{subfigure}%
\caption{(a) Photograph of MMThGEM assembly. (b) HV biasing circuit diagram. All resistors and capacitors are 200 M$\Omega$ and 10 nF respectively.}
\end{figure}

An image of the experimental setup can be seen in figure \ref{fig:MMThGEM_setup} and shows the MMThGEM mounted to an acrylic base with a standoff of 1cm and includes a cathode mounted 3cm above the top surface of the MMThGEM to form a TPC. A black 3D printed source holder was secured to the acrylic such that X-rays from an $^{55}$Fe source could be directed towards the center of the TPC volume in a repeatable way. 

 All the electrodes were biased individually with HV power supplies according to the circuit diagram shown in figure \ref{fig:MMThGEM_BIAS}. Five HV power supplies were used in the experiment; $V_c$, $V_{m1}$, $V_{m2}$, $V_{m3}$, and $V_{m4}$. These provided biasing voltages for the cathode, mesh 1, mesh 2, mesh 3 and mesh 4 respectively. $V_c$ produced a negative HV and was connected directly to the cathode. The top layer of the MMThGEM was connected to ground via a resistor to reduce the number of HV supplies required. $V_{m1-4}$ all produced positive voltages of increasing magnitude. $V_{m1}$ and $V_{m3}$ were connected to meshes 1 and 3 respectively via a resistor and low pass filter to reduce capacitative coupling between meshes. $V_{m2}$ and $V_{m4}$ were connected to meshes 2 and 4 via the biasing input on separate CREMAT CR-150 evaluation boards. The bottom electrode of the MMThGEM was left floating. All the HV channels were provided by three NHQ 202M iseg Nuclear Instrumentation Modules (NIMs). The evaluation boards utilised a CR-111 charge sensitive preamplifier whose output was connected to a CR-200-4$\mu$s shaper module on a CR-160 shaper evaluation board. The output of the shapers were connected to a NI USB-5132 8-bit 50MS/s Digitizer which interfaced with a simple labview program so that the amplified charge could be monitored on mesh 2 and measured on mesh 4. The MMThGEM assembly and biasing circuitry were placed inside a vacuum chamber and the vessel was sealed. 

\section{Calibration with Electron Drift Gas CF$_{4}$}

An energy spectrum is usually acquired by recording the amplitude of pulses from the output of the shaper. This method of acquisition is sufficient for gain measurements in electron drift gases like CF$_4$ because the charge collection time of electrons in the gas is shorter than the integration time of the shaper. Due to the NI nature of SF$_{6}$ the charge collection period of the NIs, $\sim$80 $\mu$s for measurements presented in this paper, is longer than the integration time of the shaper. This means that the pulse height of the shaper signal does not necessarily account for all the collected charge in a NI gas, therefore the integral of the shaper signal must be acquired instead. In this section we calibrate the integral of the shaper signal above a threshold of 5 mV, using the Simpsons method, against the amplitude of the shaper signal in CF$_4$ to establish a self consistent method of comparing the gas gain of an electron drift gas and a NID gas.

To begin the calibration, the air inside the vacuum chamber was evacuated using a vacuum scroll pump for 72 hours to allow sufficient time for out gassing. Before filling, the vessel was purged with CF$_4$ and evacuated for a further 10 minutes to ensure that the gas line was free from contaminants. The vessel was then filled with 40.0 $\pm$ 0.1 Torr of CF$_4$. The leak rate of the vessel was < 0.1 Torr per day. The electronics were then calibrated by injecting test pulses from a 480 Ortec Pulser NIM onto the CREMAT evaluation boards via a 1 pF capacitor. By using the amplitude of the test pulse and the capacitance, the amount of charge reaching the preamplifier was determined. The electronic gain for the amplitude and integral methods were found to be 1.39 $\pm$ 0.04 V/pC and 17.5 $\pm$ 0.5 V$\cdot\mu$s/pC respectively. The gas gain associated with this amount of charge was determined via the energy of the $^{55}$Fe X-ray (5.89 keV) and the W-value of the gas. The W-value is the same for CF$_4$ and SF$_6$ at 34 eV \cite{Reinking1986, Lopes1986}. The gas gain is defined as the amount of amplified charge reaching the preamplifier on mesh 4 divided by the initial amount of charge liberated from the gas.


Following the calibration of the electronics, the HV power supplies were ramped up to set up electric fields of varying strengths in the different detector regions. $V_{c}$, $V_{m1}$, and transfer field 1 were held constant at -300 V, 100 V, and 500 V/cm respectively. The amplification fields were increased in tandem from 18500 V/cm to 22500 V/cm in increments of 500 V/cm. These voltages were chosen as they allowed for the measurement of a broad range of gas gains and previously indicated good collection efficiency \cite{Eldridge}. For each amplification field strength, the shaper output from mesh 4 was passed to the labview program and the pulses were histogrammed according to their amplitude and integral value.

\begin{figure} [h!]
\captionsetup[subfigure]{justification=centering, labelformat=empty}
\centering
\begin{subfigure} {.5\textwidth}
    \centering
  \includegraphics[width=\textwidth]{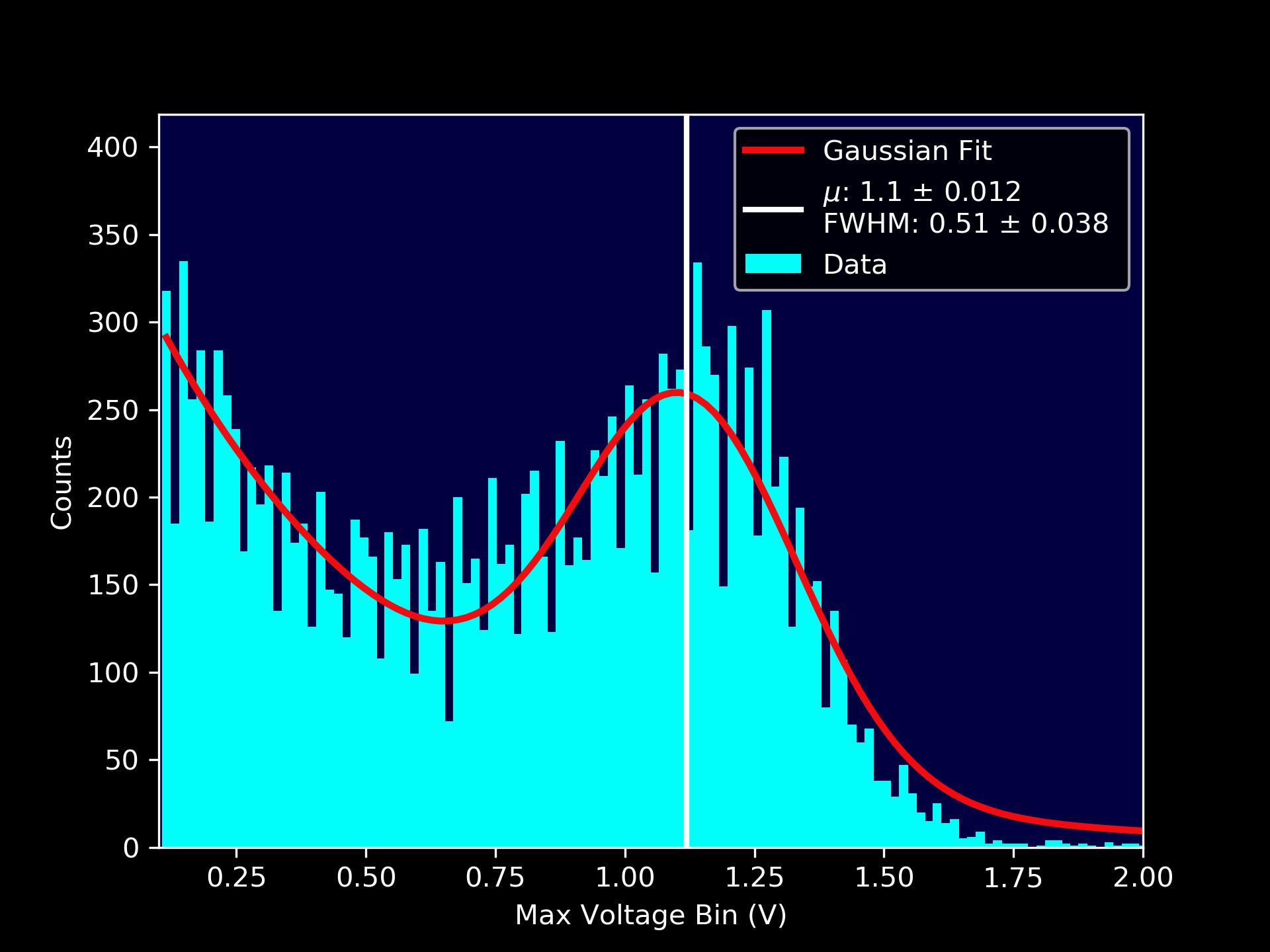}
  \caption{}
  \label{fig:ex_max_spec}
\end{subfigure}%
\begin{subfigure} {.5\textwidth}
  \centering
  \includegraphics[width=\textwidth]{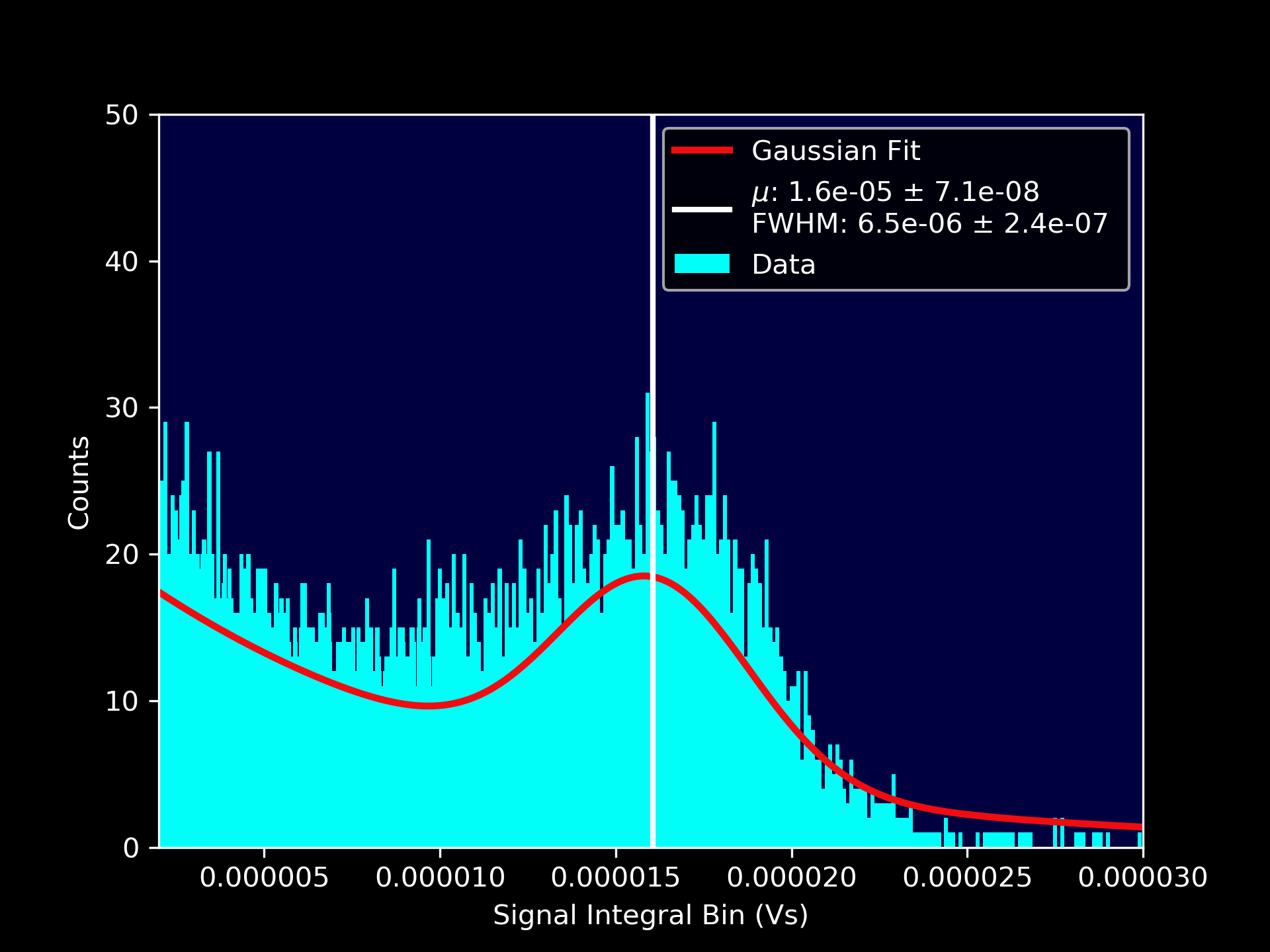}
  \caption{}
  \label{fig:ex_int_spec}
\end{subfigure}%
\vspace{-1.4\baselineskip}
\caption{Shaper amplitude spectrum with both amplification fields set to 21000 V/cm in 40 Torr CF$_4$ (left). Shaper signal integral spectrum with both amplification fields set to 21000 V/cm in 40 Torr CF$_4$ (right).}
\label{fig:spec}
\end{figure}

After a 30 minute exposure, spectra could be observed in the amplitude and integrated signal histograms, examples of which can be seen in figure \ref{fig:spec}. These spectra exhibit a clear photopeak, caused by the complete absorption of the $^{55}$Fe X-ray, and an exponentially falling component; caused by partial/inefficient collection, transfer and amplification of the initial charge. The spectra were then fitted with both an exponential curve and a Gaussian to the photopeak, shown in red. The mean of the Gaussian was then determined, indicated by the white vertical line, and the gas gain calibrations were applied to the mean of the Gaussian curves. 

\begin{figure} [h!]
\centering
\includegraphics[trim ={2.0cm 0.4cm 2.0cm 0.4cm}, clip, width=\textwidth]{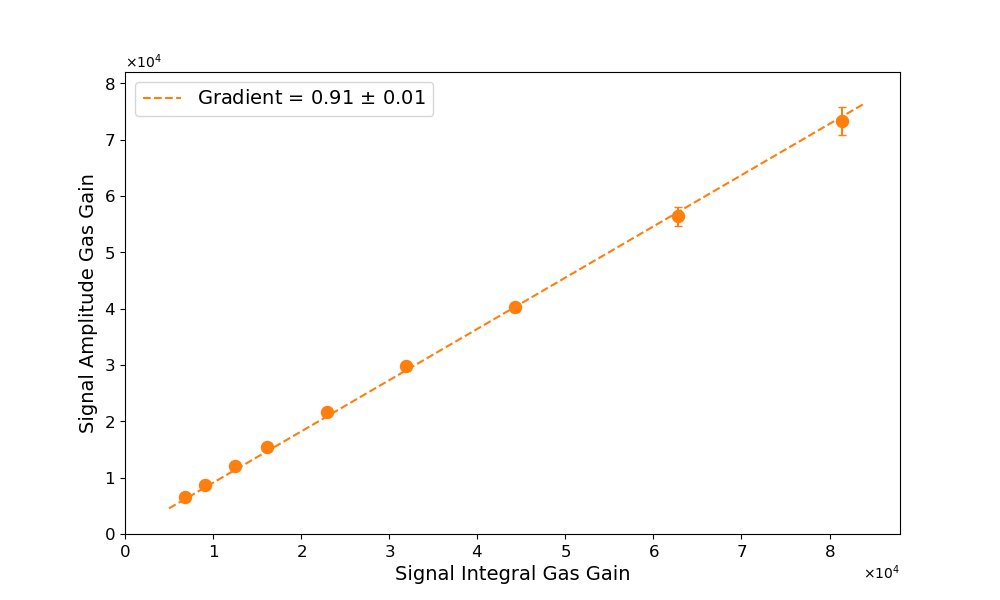}
\caption{Gas gain measurements according to the amplitude method vs those of the integrated signal method in 40 Torr CF$_4$.}
\label{fig:integral_signal}
\end{figure}

The resulting gas gains from both the amplitude and integral methods can be seen plotted against each other in figure \ref{fig:integral_signal}. A linear regression analysis was performed with the intercept passing through (0, 0). The gradient of this line should be equal to unity, however it was found that the gradient was 0.91 $\pm$ 0.01. This slight discrepancy is likely an artifact of the integral method. For the purpose of establishing a self consistent method of comparing the gas gains in CF$_{4}$ to that of SF$_{6}$, all further gas gain measurements using the integral method were re-calibrated against the signal amplitude method according to figure \ref{fig:integral_signal}.

\section{Gas Gains in SF$_6$ with the MMThGEM}
\label{sec:pre-optimisation}

Previous work conducted with the MMThGEM in 20 and 30 Torr of SF$_6$ has shown that gas gains were limited by a ringing effect at larger amplification field strengths. This caused the shaper signal to oscillate and was unable to return to its baseline without intervention. However, this ringing effect was not observed at 40 Torr as the device was not pushed to its physical limitations out of caution \cite{Eldridge}. In this section, the MMThGEM is pushed to its sparking limit and then an optimisation procedure is performed on both the collection and transfer field 1 to explore the maximum achievable gas gain in 40 Torr of SF$_6$.

\begin{figure} [h]
\centering
    \includegraphics[trim ={3cm 0.5cm 3cm 2cm}, clip, width = \textwidth]{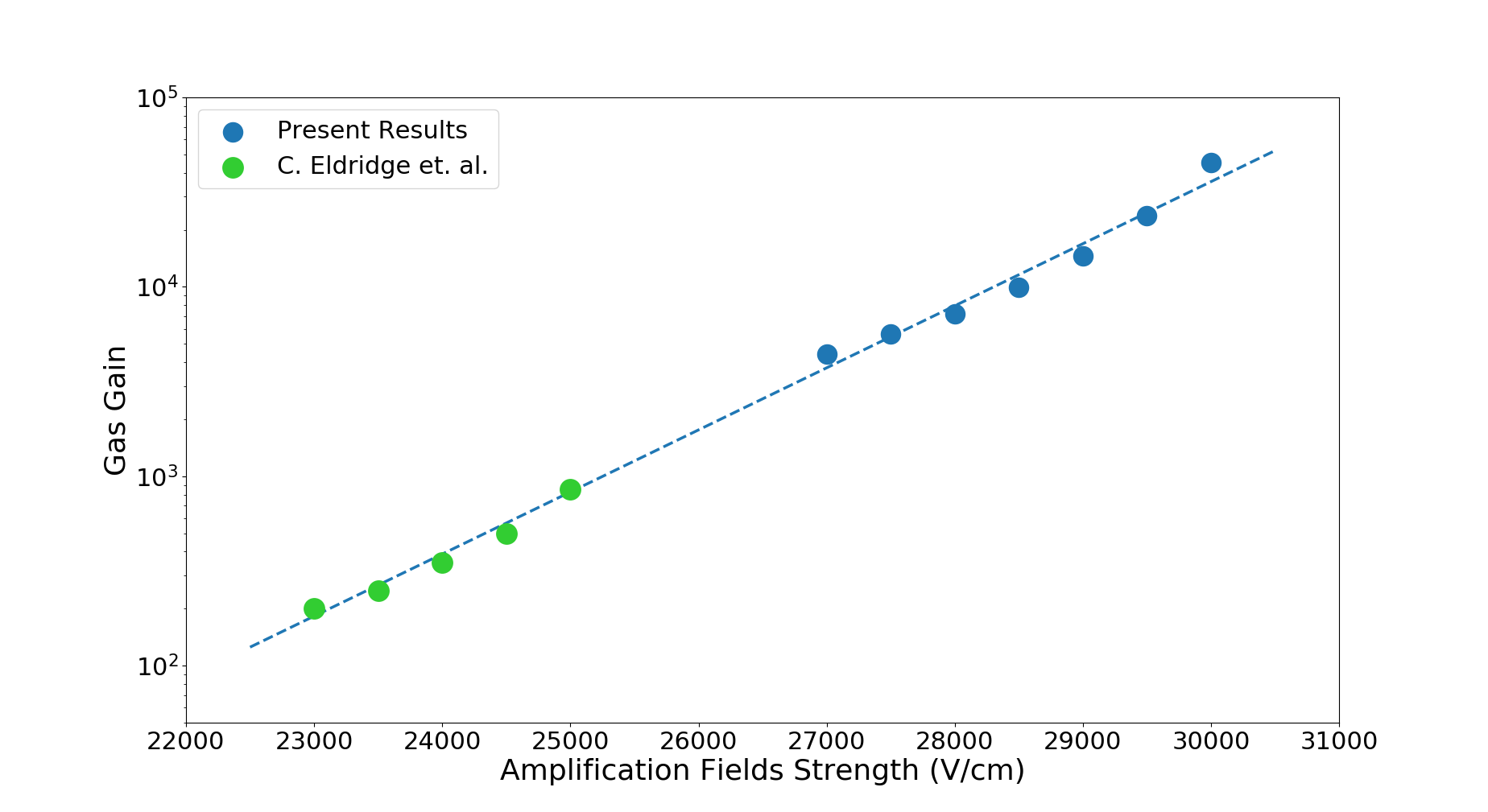}
    \caption{Gas gain vs amplification fields strength in 40 Torr SF$_6$. Error bars are smaller than the marker size and therefore not observed.}
  \label{fig:callum_comparison}
\end{figure}

To begin, the vessel was filled with SF$_6$ in an identical manner to that of the CF$_4$ measurements. $V_c$, $V_{m1}$, and transfer field 1 were then set to -500 V, 30 V, and 600 V/cm respectively. The drift field was chosen to give good collection efficiency while the collection and transfer settings were chosen because they were previously used in 40 Torr of SF$_6$ \cite{Eldridge}. The amplification fields were then increased in tandem until a photopeak could be observed on the integrated signal spectrum, which occurred above 27000 V/cm. Integral spectra were acquired and subsequent gas gains were calculated for amplification field strengths between 27000 and 30000 V/cm in increments of 500 V/cm. Above 30000 V/cm sparking and ringing events which did not return to baseline were observed. The gas gain measurements were plotted as a function of the amplification field strengths and can be seen in blue in figure \ref{fig:callum_comparison}. As expected, the gas gain measurements exhibit an exponential increase with amplification fields strength. The gas gains achieved range from 4400 $\pm$ 400 to a maximum of 45200 $\pm$ 500. This maximum gas gain is an order of magnitude larger than what has been observed previously in SF$_6$ with the MMThGEM \cite{Eldridge} and other MPGD devices \cite{Phan2017, Baracchini2018, Miyamoto2004}. Moreover this is comparable to the gas gains achieved in the CF$_4$ calibration run. 

Previous results in 40 Torr of SF$_6$, as presented in \cite{Eldridge}, have also been included in figure \ref{fig:callum_comparison} and can be seen in green. The exponential curve which was fitted to the present results was extrapolated down to these lower field strengths. It can be seen that the fitted line appears to have good agreement with the previously obtained results. To quantify this agreement, an R$^2$ test was performed between the extrapolated fit and previous measurements which produced a value of 0.97. This result provides strong evidence that the work presented in this paper is consistent with previously obtained results. Direct measurement at these lower amplification fields could not be made because the dynamic range of the digitizer used in these measurements was not sufficient to see these smaller signals above the electronic noise.

\begin{figure} [t]
\centering
    \includegraphics[trim ={3cm 0.9cm 3cm 2.2cm}, clip, width = \textwidth]{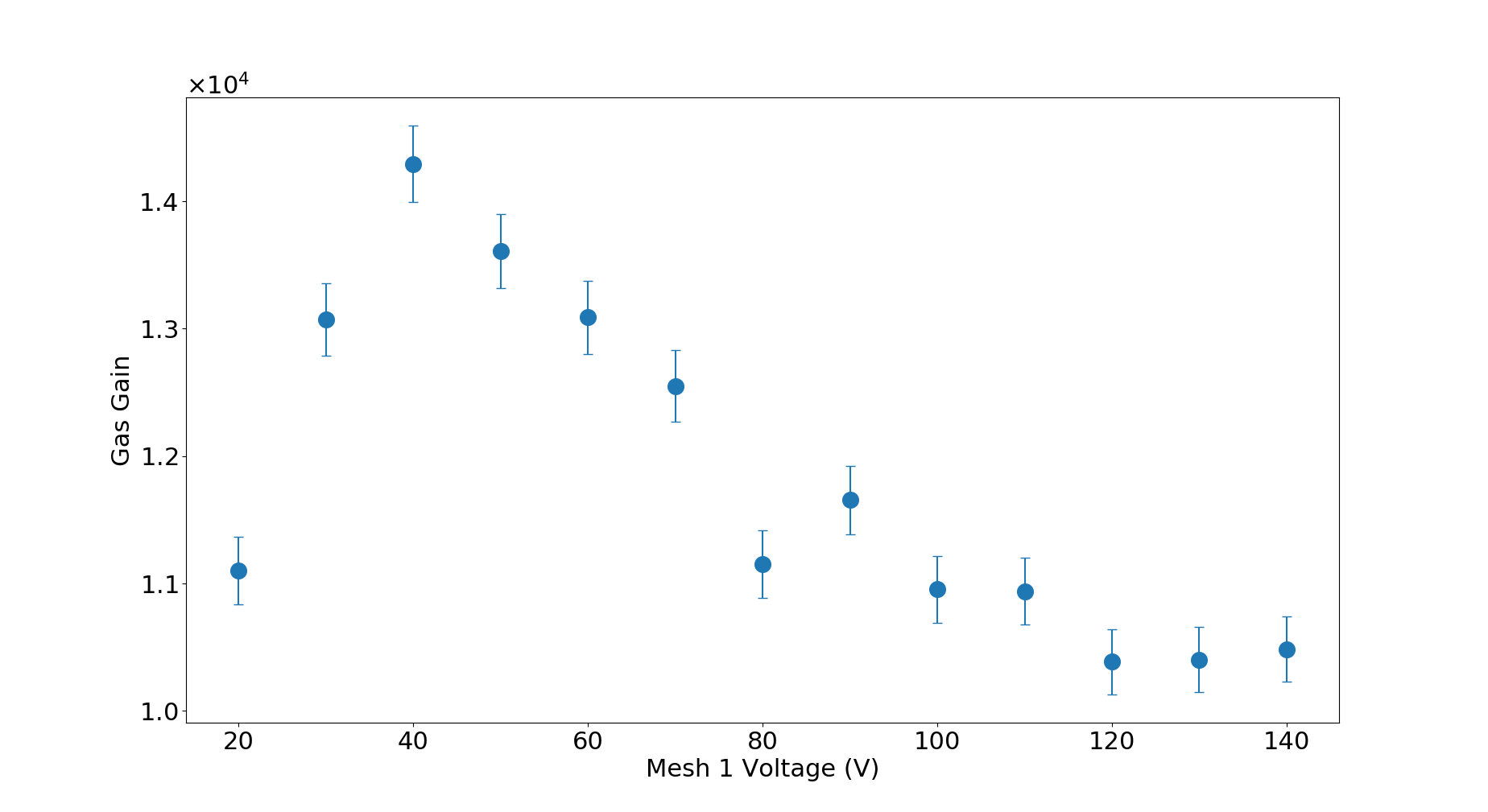}
    \caption{Gas gain vs $V_{m1}$ showing optimum collection voltage at 40 V in 40 Torr SF$_6$.}
  \label{fig:collection_opt}
\end{figure}

Although this gas gain result is a significant improvement on previous results with SF$_6$, this is not a thoroughly optimised detector regime. To push the device to its physical limits in SF$_6$, the collection and transfer fields underwent methodical optimisation. To optimise the collection field; $V_c$, the amplification and transfer fields were held constant at -500 V, 29000 V/cm and 600 V/cm respectively as this produced a mid range gas gain in figure \ref{fig:callum_comparison}. The collection field was varied by increasing $V_{m1}$ from 20 V to 140 V in increments of 10 V. Figure \ref{fig:collection_opt} shows the resulting gas gains plotted against V$_{m1}$. It can be seen that, as $V_{m1}$ increased to 40 V, the gas gain increased from 11100 $ \pm$ 300 to 14300 $\pm$ 300. Above 40 V the gas gain began to decrease slowly down to 10500 $\pm$ 300 at 140 V. The clear peak in gas gain observed at 40 V was determined to be the optimal collection field strength for maximising gas gain.

\begin{figure} [b]
\centering
    \includegraphics[trim ={3cm 1cm 3cm 2.2cm}, clip, width = \textwidth]{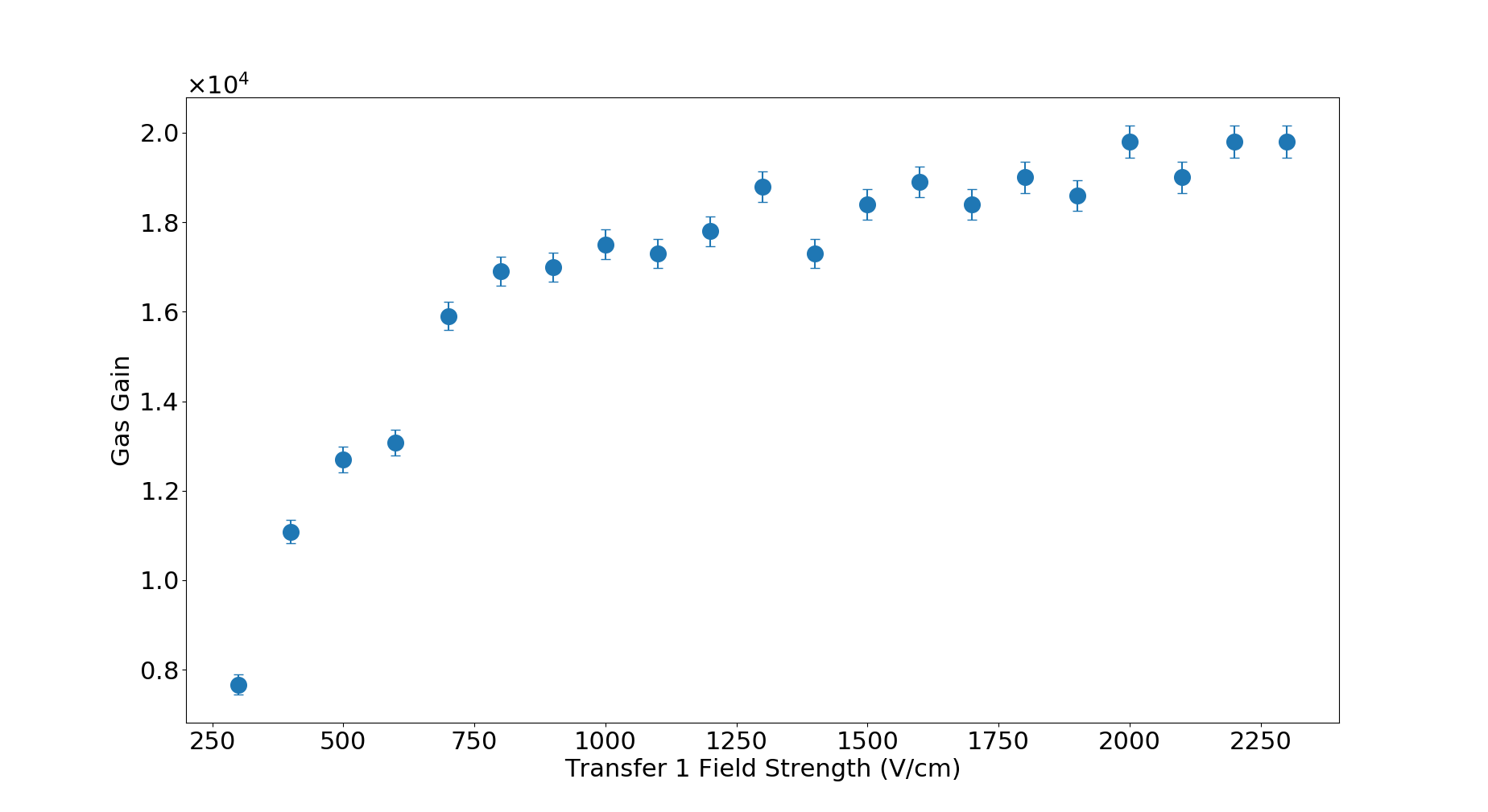}
    \caption{Gas gain vs transfer 1 field strength exhibiting a plateau above 900 V/cm in 40 Torr SF$_6$.}
  \label{fig:transfer_opt}
\end{figure}

Now that the optimum collection field has been determined, the transfer field strength can be subjected to a similar gain optimisation procedure. This was achieved by holding $V_c$, $V_{m1}$, and the amplification fields constant at -500 V, 30 V, and 29000 V/cm respectively. The transfer field was then increased from 300 to 2300 V/cm in increments of 100 V/cm and resulting gas gains were determined. These results can be seen in figure \ref{fig:transfer_opt} and show that, as the transfer field increases from 300 to 900 V/cm, the gas gain increases from 7700 $\pm$ 200 to 17000 $\pm$ 300. Above 900 V/cm the gas gain begins to plateau and the improvement in gas gain becomes less significant with increasing strength. Therefore increasing the transfer field strength beyond the plateau was not beneficial for performance and thus 900 V/cm was taken as the optimised transfer field strength.

\begin{figure} [b]
\centering
    \includegraphics[trim ={2cm 0.4cm 2cm 0cm}, clip, width = \textwidth]{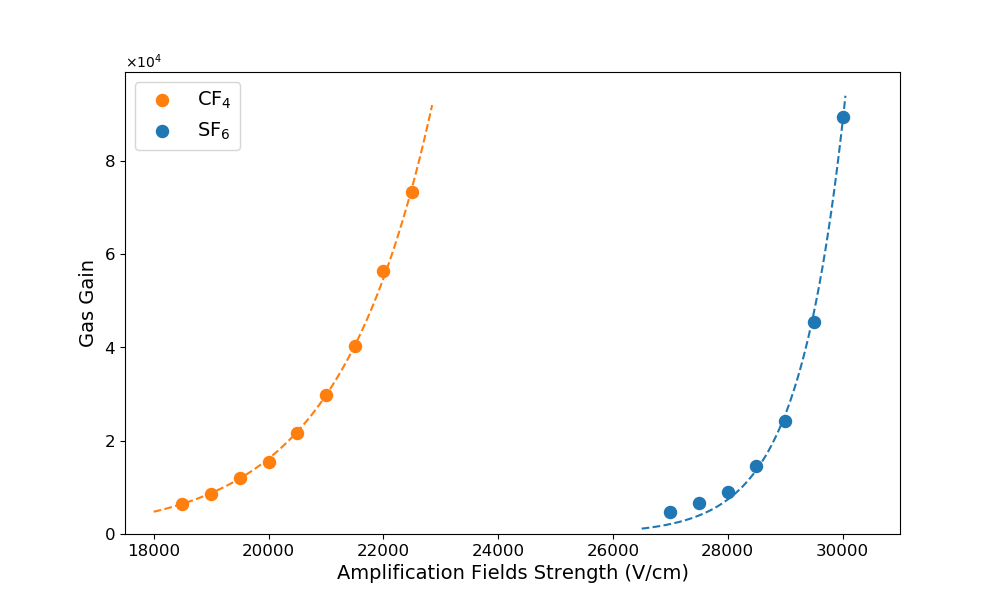}
    \caption{Gas gain vs amplification field strengths in 40 Torr CF$_4$ and fully optimised in 40 Torr SF$_6$.}
  \label{fig:fully_opt}
\end{figure}

Once the optimised field strengths for both the collection and transfer fields were determined, the gas gain of the fully optimised device could be tested. By holding $V_c$, $V_{m1}$, and the transfer field strength constant at -500 V, 40 V and 900 V/cm, the amplification fields were once again increased in tandem from 27000 to 30000 V/cm in increments of 500 V/cm. The gas gain measurements that resulted can be seen in figure \ref{fig:fully_opt}. 

Figure \ref{fig:fully_opt} shows that as the amplification fields increase from 27000 to 30000 V/cm, the gas gain increases exponentially from 4700 $\pm$ 200 to a maximum of 89300 $\pm$ 600 before sparking and continuous ringing was observed at higher field strengths. The optimisation process constitutes an improvement in gas gain by a factor of $\sim$ 2 compared to the maximum gain achieved in figure \ref{fig:callum_comparison} prior to the optimisation. This demonstrates that the gas gain is strongly dependent on the field strengths of all the regions in the MMThGEM, not just the amplification field strengths. It is worth mentioning that a minority of ringing events were observed during the 29500 and 30000 V/cm measurements, however these operating voltages were stable as the signal voltage was able to return to baseline without intervention and these signals did not account for a significant portion of events. Further investigation is required to fully understand the suppression of this ringing phenomenon.

The gas gains achieved with the fully optimised device are comparable to the readily attainable gas gains achieved during the CF$_4$ calibration run. However, the amplification fields required were significantly larger in SF$_6$. While CF$_4$ could produce gas gains on the order of $10^4$ above 19000 V/cm, SF$_6$ required amplification field strengths greater than 28000 V/cm. This large increase of 9000 V/cm is unsurprising for SF$_6$ due to the electronegative nature of the gas. Despite the much larger field strengths required, these results show significant promise for the MMThGEM as an amplification stage in the next generation of NITPC directional DM searches. The order of magnitude improvement demonstrated throughout this paper will ultimately benefit the sensitivity of these detectors to low energy NRs. 

\begin{figure} [b]
    \centering    \includegraphics[trim ={1.5cm 0.4cm 1cm 1.35cm}, clip, width = \textwidth]{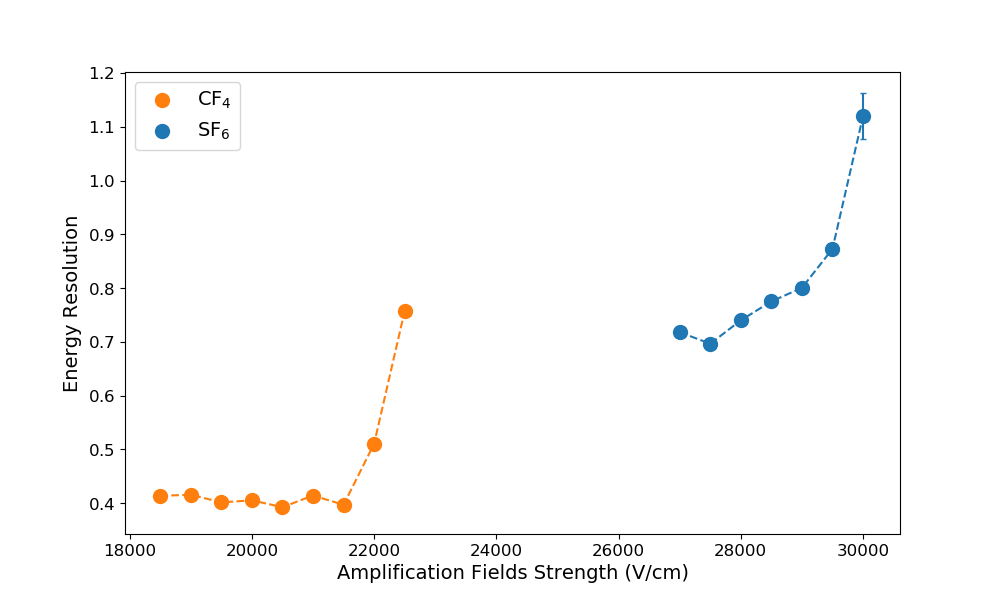}
    \caption{Energy resolution vs amplification fields strength in 40 Torr CF$_4$ and 40 Torr SF$_6$ after the gain optimisation.}
    \label{fig:ER}
\end{figure}

For completeness, a comparison of the energy resolution between the CF$_4$ and SF$_6$ measurements has been included in figure \ref{fig:ER}. It can be seen that the energy resolution in CF$_4$ is generally better than that of SF$_6$. As the amplification fields increase from 18500 to 21500 V/cm, the energy resolution mostly decreases from 0.4136 $\pm$ 0.0008 to 0.397 $\pm$ 0.002 with the smallest energy resolution of 0.392 $\pm$ 0.001 occurring at 20500 V/cm. It then increases significantly to 0.758 $\pm$ 0.005 at 22500 V/cm. The best energy resolution observed in SF$_6$ is comparable to this. At 27000 V/cm the energy resolution starts at 0.7180 $\pm$ 0.0005 before dropping slightly to its lowest of 0.696 $\pm$ 0.001 at 27500 V/cm. The energy resolution then becomes progressively worse up to 1.12 $\pm$ 0.04 at 30000 V/cm. Although the gain in SF$_6$ is excellent at 30000 V/cm, the energy resolution is poor and could possible benefit from a similar optimisation procedure in future. \hfill

\section{Conclusion}
\label{sec:conclusions}

In conclusion, the integral method required for measuring the gas gain of SF$_6$ was first calibrated against the standard amplitude method in CF$_4$. Then the MMThGEM was pushed to its sparking limit in 40 Torr of SF$_6$. These initial results were able to produce a maximum gas gain of 45200 $\pm$ 500, which was larger than previously demonstrated results by an order of magnitude. The collection and transfer field 1 were then isolated and subjected to an optimisation process. The device was found to produce optimum gas gains when $V_{m1}$ and transfer field 1 strengths were set to 40 V and 900 V/cm respectively. The optimised field settings were then tested by holding $V_c$, $V_{m1}$, and transfer field 1 constant while increasing both the amplification fields in tandem. This yielded an absolute maximum gas gain of 89300 $\pm$ 600 before sparking occurred. These results are significant because they demonstrate gas gains in a NID gas greater than 10$^4$ for the first time. This is at least an order of magnitude improvement on what was previously considered possible with the gas at this low pressure and ultimately offers a significant advancement in the sensitivity of NITPCs to low energy recoils in the context of a directional dark matter search. The energy resolution was also evaluated in both CF$_4$ and fully optimised SF$_6$. It was found that the energy resolution was generally better in CF$_4$ with the lowest value reaching 0.392 $\pm$ 0.001 compared to 0.696 $\pm$ 0.001 in SF$_6$. The energy resolution could benefit from a similar optimisation procedure in future. Finally, it is recommended that future MMThGEM designs should focus on reducing the pitch and diameter of the hole structures to aid positional resolution and track reconstruction when coupled to a micromegas.

\acknowledgments
The authors would like to acknowledge support for this work from AWE, the CERN MPGD group and a University PhD scholarship awarded to A.G. McLean.






\end{document}